\documentclass[6pt,twocolumn,showpacs,preprintnumbers,amsmath,amssymb,pra]{revtex4}

\usepackage{color}
\usepackage{graphicx}
\usepackage{dcolumn}
\usepackage{bm}
\usepackage{cancel}
\usepackage{soul}

\newcommand{\dint}[2]{\mathrm{d}^{#1}{#2}}

\begin{document}


\title{Efficient grid-based method in nonequilibrium Green's function calculations.\\
Application to model atoms and molecules}

\author{K. Balzer}
\affiliation{Institut f\"ur Theoretische Physik und Astrophysik, Christian-Albrechts-Universit\"at Kiel, Leibnizstrasse 15, 24098 Kiel, Germany}
\email{balzer@theo-physik.uni-kiel.de}
\author{S. Bauch}
\affiliation{Institut f\"ur Theoretische Physik und Astrophysik, Christian-Albrechts-Universit\"at Kiel, Leibnizstrasse 15, 24098 Kiel, Germany}
\author{M. Bonitz}{
\affiliation{Institut f\"ur Theoretische Physik und Astrophysik, Christian-Albrechts-Universit\"at Kiel, Leibnizstrasse 15, 24098 Kiel, Germany}
\date{\today}

\begin{abstract}
We propose and apply the finite-element discrete variable representation to express the nonequilibrium Green's function for strongly inhomogeneous quantum systems. This method is highly favorable against a general basis approach with regard to numerical complexity, memory resources, and computation time. Its flexibility also allows for an accurate representation of spatially extended hamiltonians, and thus opens the way towards a direct solution of the two-time Schwinger/Keldysh/Kadanoff-Baym equations on spatial grids, including e.g.~the description of highly excited states in atoms. As first benchmarks, we compute and characterize, in Hartree-Fock and second Born approximation, the ground states of the He atom, the H$_2$ molecule  and the LiH molecule in one spatial dimension. Thereby, the ground-state/binding energies, densities and bond-lengths are compared with the direct solution of the time-dependent Schr\"odinger equation.
\end{abstract}

\pacs{02.70.Dh, 05.30.-d, 31.15.-p}


\maketitle

\section{\label{sec:intro}Introduction}
The two-time Schwinger/Keldysh/Kadanoff-Baym equations (SKKBE), e.g.~\cite{martin59,keldysh64,kadanoff62}, allow for a quantum statistical analysis of nonequilibrium processes on microscopic footing. To great success, the one-particle nonequilibrium Green's function (NEGF) has been computed from the SKKBE for a variety of homogeneous quantum systems, e.g.~for nuclear matter\cite{danielewicz84,koehler95,bozek97}, the correlated electron gas\cite{kwong00}, dense plasmas\cite{bonitz96,semkat99,semkat00}, or electron-hole plasmas\cite{binder97,kwong98,schaefer,banyai,jahnke}, where different types of many-body approximations, by diagram technique, have been included in a conserving manner. On the contrary, NEGFs, only in the recent decade, have challenged attention with respect to spatial inhomogeneity, exploring localized, finite and strongly correlated systems. Examples are electrons in atoms and small molecules\cite{stan06,dahlen07_prl}, few-electron quantum dots\cite{balzer09_prb,balzer09_jpa} and charge carriers in lattice and transport models such as strongly correlated Hubbard chains\cite{vanfriesen09}, molecular junctions\cite{thygesen08} and quantum dot levels coupled to leads\cite{myohanen08}.

Although computational capabilities have been permanently increasing in the recent past, NEGF calculations remain a demanding task for finite systems:~in particular---including electron-electron correlations in second Born approximation---highly excited states in atoms or time-dependent phenomena related to their ionization are generally difficult to access, and, only very few attempts have been made\cite{hochstuhl09, bonitz09}. Also, the describability of specific correlation effects, such as two-electron resonances in dipole spectra\cite{tanner00}, remain so far unanswered in NEGF approaches as they require an accurate and extensive (large-scale) computation of the temporal evolution following an intense external perturbation.

All above-mentioned finite systems rely on general \mbox{(semi-)analytic} basis expansions of the nonequilibrium Green's function. Nevertheless, concerning the numerical complexity associated with the NEGF, a basis representation reveals restricted capabilities. This affects, in particular, the spatial resolution, where a relatively small number of
single-particle orbitals (typically $n_b\lesssim60$ are feasible) are not appropriate to resolve the nonequilibrium dynamics when, e.g.~in atoms, occupations of highly excited states are non-negligible or ionization processes are involved. The same is the case when specialized basis sets constructed from Gauss-type or Slater-type orbitals or potential eigenstates are being used. To this end, extremely large basis sets are needed and the system under investigation requires a large-scale treatment. 

Another option is provided by grid-based methods. However, for inhomogeneous systems, no direct solution of the SKKBE with grid-based---and, in turn, finite-difference---methodologies has been performed so far, that systematically includes binary correlations and memory effects. This is due to the fact that numerical grid methods allow for intuitive control but require small mesh spacings which become impractical for the compound structure of the two-space two-time Green's function: We note, that, in full three space dimensions (3D), the NEGF is an eight-dimensional complex function. However, also in one spatial dimension, generally severe problems arise in the framework of spatially extended hamiltonians, where particles may occupy broad domains in coordinate and/or momentum space. Thus, an alternative method for NEGF calculations is desirable, to which one attributes more numerical flexibility and efficiency, and which has the ability to combine the advantages of non-existent grid and standard basis approaches.

\begin{figure*}[t]
 \includegraphics[width=32.5pc]{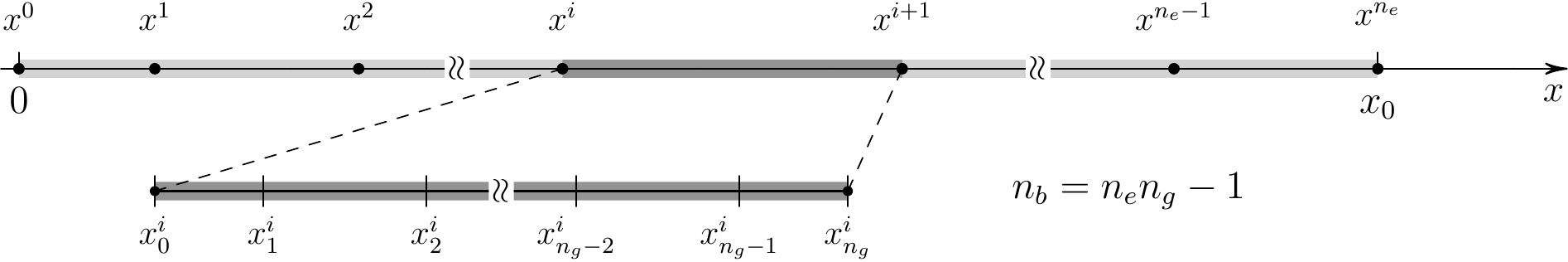}
 \caption{
In FE-DVR representation, the interval $[0,x_0]$ is partitioned into $n_e$ finite elements $[x^i,x^{i+1}]$. In each FE, $n_g$ generalized Gauss-Lobatto points (denoted $x_m^i$) provide the basis for the construction of a local DVR basis set. $n_b$ denotes the dimensionality of the extended basis covering the whole interval.
}\label{fig1}
\end{figure*}

In this paper, we develop such a computational method based on the finite-element discrete variable representation (FE-DVR), see Refs.~\cite{rescigno00,schneider06} and Sec.~\ref{subsec:basisconstruction}. This method allows for an efficient solution of the two-time SKKBE for the one-particle Green's function, at least, in one spatial dimension. As general system, we, thereby, consider $N$ interacting electrons, the non-relativistic hamiltonian of which reads
\begin{align}
\label{ham}
 \hat{h}&=\hat{t}+\hat{v}+\hat{u}&\\
&=-\frac{1}{2}\sum_{i=1}^{N}\nabla^2_i+\sum_{i=1}^{N}v(x_i,t)+\sum_{i<j}u(|x_i-x_j|)\;,\nonumber
\end{align}
with kinetic energy $\hat{t}$, a possibly time-dependent potential energy $\hat{v}$, and the binary interactions described by $\hat{u}$. Except for their spin orientations, all electrons are considered identical (in mass and charge), and, throughout the present work, we will use atomic units.

The use of the FE-DVR provides analytical expressions for the kinetic and potential energy in NEGF calculations for strongly inhomogeneous systems. But the main achievement of the present paper is the first realization of a grid-based NEGF approach \textit{together} with a very efficient treatment of the binary interactions. Explicitly, instead of ${\cal O}(n_b^4)$ interaction matrix elements, see Sec.~\ref{subsec:matrixelements}, our method requires only ${\cal O}(n_b^2)$ elements, which, in addition, need not to be precomputed as before in a complicated manner. With regard to the SKKBE, the latter point directly leads to much simpler, semi-analytical formulas for the first- and second-order self-energies, which are independent of the explicit form of the interaction, Sec.~\ref{subsec:skkbe}. With these remarkable scaling properties, the FE-DVR essentially reduces the numerical effort, such that considerably less storage memory and computing time is needed, and, hence, enables calculations on significantly larger, more extended systems than before.

In Sec.~\ref{sec:results}, we demonstrate the power of the approach and compute the nonequilibrium Green's functions for the one-dimensional He atom and the neutral molecules H$_2$ and LiH (also in one spatial dimension) as function of the interatomic distance. In the course of this,
we focus on the ground-state properties and compare the Hartree-Fock and the second Born approximation to the exact solution obtained from the full few-particle Schr\"odinger equation. Ignoring the nuclear dynamics [i.e.~in the Born-Oppenheimer scheme], the exploration of nonequilibrium properties is straightforward within the formalism presented. However, a detailed discussion is deferred to a forthcoming publication.

\section{\label{sec:fedvr}Finite-element discrete variable representation}

The finite-element discrete variable representation is a hybrid approach\cite{collins04} which combines finite-element (FE) methods, i.e.~spatial grids, and the discrete variable representation\cite{light85} (DVR). In a DVR basis, a similarity transformation allows us to replace matrix elements of local operators [of the coordinates] by their values on a relatively small numerical grid. The high degree of accuracy of this procedure, widely used in quantum chemistry, manifests its usefulness in solving quantum mechanical problems\cite{light07}.

For the direct solution of the few-particle time-dependent Schr\"odinger equation, e.g.~Ref.~\cite{feist09} and references therein, the FE-DVR is highly effective---often in combination with time-dependent close coupling (TDCC)---due to the accuracy of the DVR on the one hand and the sparse character of FEs on the other. However, these scaling properties, that enable a well parallelizable code\cite{schneider06}, are less important for our application of the method. Instead, we focus on the benefits of the FE-DVR regarding the treatment of binary interactions and self-energies, which require the main computational expense within the framework of nonequilibrium Green's functions.

The general idea how to combine FEs with the DVR to construct an extended basis is outlined in the following. Thereafter, in Sec.~\ref{subsec:matrixelements}, we discuss and give formulas for the relevant matrix elements of $\hat{t}$, $\hat{v}$ and $\hat{u}$, which are finally used in the equations of motions for the one-particle Green's function, see Sec.~\ref{subsec:skkbe}.

\subsection{\label{subsec:basisconstruction}Basis construction}
We divide the interval $[0,x_0]$, which is of physical and numerical relevance regarding hamiltonian (\ref{ham}) and may be spatially extended, into $n_e$ finite elements with arbitrary boundaries $x^0=0<x^1<x^2<\ldots<x^{n_e-1},x^{n_e}=x_0$, see Fig.~\ref{fig1}. In each FE $i$, i.e.~in $[x^i,x^{i+1}]$, we then construct a local DVR basis based on the generalized Gauss-Lobatto points\cite{rescigno00} $x^i_m$ and weights $w_m^i$:
\begin{align}
 x^i_m&=\frac{1}{2}\left\{\left(x^{i+1}-x^i\right)x_m + \left(x^{i+1}+x^i\right)\right\}\;,\nonumber\\
 w^i_m&=\frac{w_m}{2}\left(x^{i+1}-x^i\right)\;.
\end{align}
When using $n_g$ Legendre interpolating functions, the points $x_m$ (\textit{standard} Gauss-Lobatto points) are defined as roots of the first derivative of Legendre polynomials $P_n(x)$ according to
\begin{align}
 \frac{\mathrm{d}}{\mathrm{d}x}P_{n_g}(x_m)=0\;, 
\end{align}
and the associated weights are
\begin{align}
w_m=\displaystyle\frac{2}{n_g(n_g+1)[P_{n_g}(x_m)]^2}\;.
\end{align}

\begin{figure}[t]
 \includegraphics[width=0.485\textwidth]{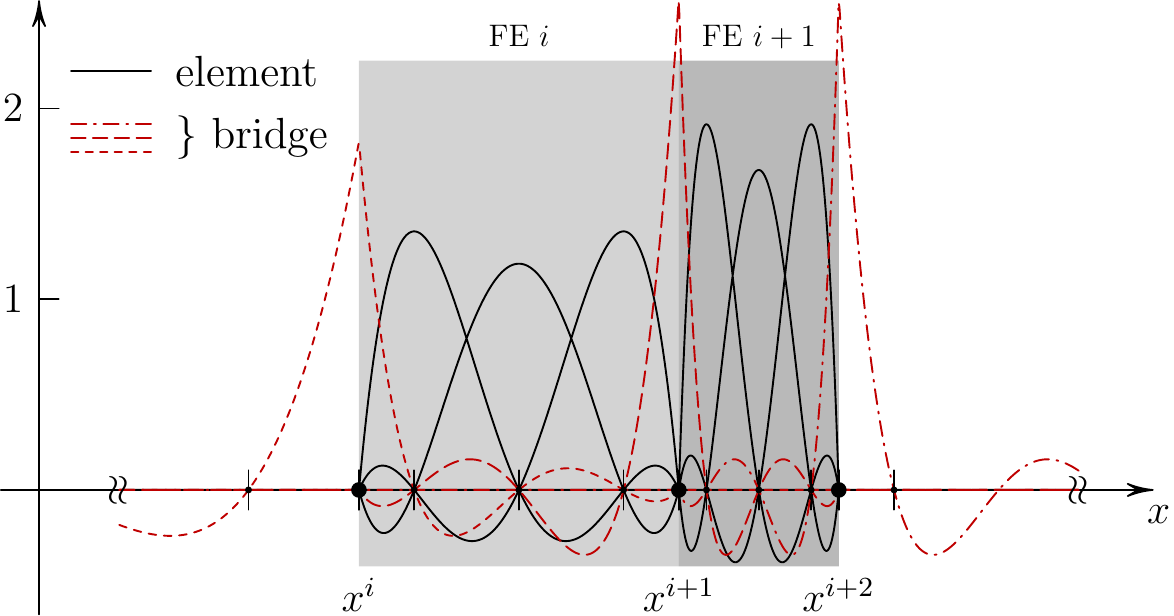}
 \caption{(Color online) Structure of a FE-DVR basis $\{\chi^i_m(x)\}$ with $n_g=4$, i.e. $5$ local DVR basis functions in each element. While the 'element' functions (solid) are defined in a single FE, the 'bridge' functions  (dashed and dash-dotted lines) link two adjacent FEs.}\label{fig1a}
\end{figure}

In our approach, we use a DVR basis of equal size in each FE, see Fig.~\ref{fig1a}. The generalization to different numbers of basis functions per element is straightforward, and only slightly alters the matrix elements involved, cf.~Sec.~\ref{subsec:matrixelements}. The one-dimensional FE-DVR space is spanned by the set of orthonormal\cite{gglorthonormal} functions
\begin{align}
\label{basis}
 \chi_{m}^{i}(x)&=
\left\{
\begin{array}{cc}
 \displaystyle\frac{f_{n_g-1}^{i}(x)\,+\,f_{0}^{i+1}(x)}{\sqrt{w^i_{n_g-1}+w^{i+1}_0}}&\;\;,\;m=0\hspace{1pc}\textup{(bridge)}\\
&\\
 \displaystyle\frac{f_{m}^{i}(x)}{\sqrt{w^i_m}} & ,\;\textup{else\hspace{1pc}(element)}
\end{array}
\right.\;,
\end{align}
with Lobatto shape functions\cite{manolopoulos88, rescigno00}
\begin{align}
f_{m}^{i}(x)=\prod_{\bar{m}\neq m} \displaystyle\frac{x-x_{\bar{m}}^i}{x_m^i-x_{\bar{m}}^i}
\end{align}
for $x^i\leq x\leq x^{i+1}$ and $f^i_m(x)=0$ for $x<x^i$ as well as $x>x^{i+1}$, which have the property $f^i_m(x_{m'}^{i'})=\delta_{ii'}\delta_{mm'}$ and are orthogonal with regard to the generalized Gauss-Lobatto quadrature, see Appendix. The 'bridge' function ($m=0$) in Eq.~(\ref{basis}) extends over two adjacent elements [element $i$ has overlap with element $i+1$] and, hence, ensures \textit{communication} between different grid domains $i$ and $i'$ and guarantees continuity of any expanded quantity or Green's function, cf.~Sec.~\ref{subsec:skkbe}. The 'element' functions are zero at and outside the element boundaries. Generally, in Eq.~(\ref{basis}) and in the remainder of this paper, superscripts are labeling elements $i$ ranging $0,1,2,\ldots,n_e-1$, and subscripts are denoting local DVR indices $m$ ranging $0,1,2,\ldots,n_g-1$, compare with Fig.~\ref{fig1}. In the first (last) FE, the DVR basis function that is part of the left-(right-)hand bridge is removed, assuming the many-body wave function of system (\ref{ham}) to vanish outside the interval $[0,x_0]$. Hence, the total basis set has dimension
\begin{align}
\label{basisdim}
 n_b=n_e n_g-1\;.
\end{align}
We note, that, with our construction of the spatial grid, see Fig.~\ref{fig1}, the formula for the dimensionality slightly differs from Refs.~\cite{rescigno00,schneider06}. Here, we do not, separately, define the size of the local DVR basis set, which would be $n_g+1$, compare with Fig.~\ref{fig1a}. Moreover, a generalization to higher dimensions is possible by using a product ansatz for the coordinate functions\cite{schneider06}.

\subsection{\label{subsec:matrixelements}Matrix elements of operators $\hat{t}$, $\hat{v}$, and $\hat{u}$}
To perform NEGF calculations with respect to system (\ref{ham}), we need the matrix elements associated with the kinetic, potential and interaction energy operators referring to the chosen FE-DVR basis. Thereby, integrations over coordinate space are calculated by using the generalized Gauss-Lobatto (GGL) quadrature, and case differentiations arise from the fact that the basis functions $\chi_m^i(x)$ split into element and bridge functions.

The potential-energy matrix---and the matrix of any other local operator---turns out to be diagonal with regard to elements $i$ and local DVR basis indices $m$:
\begin{align}
\label{vmat}
 v_{m_1m_2}^{i_1i_2}(t)&=\int_{0}^{x_0}\!\!\!\dint{}{x}\,\chi_{m_1}^{i_1}(x)\,v(x,t)\,\chi_{m_2}^{i_2}(x)&\nonumber\\
                       &=\delta_{i_1i_2}\,\delta_{m_1m_2}\,\tilde{v}_{m_1}^{i_1}(t)\;,
\end{align}
with
\begin{align}
 \tilde{v}_{m}^{i}(t)=
\left\{
\begin{array}{cc}
 v(x^{i}_{m},t) &,\;m>0\\
&\\
\displaystyle\frac{v(x_{n_g-1}^{i},t)\,w_{n_g-1}^{i}+v(x_{0}^{i+1},t)\,w_{0}^{i+1}}{w_{n_g-1}^{i}+w_{i+1}^{0}} &,\;m=0
\end{array}
\right.\;.
\end{align}
Hence, Eq.~(\ref{vmat}) implies that the potential energy is simply represented by a vector of dimension $n_b$.

The operator of the kinetic energy is non-local as it involves information of different points in physical space. As a consequence, $t_{m_1m_2}^{i_1i_2}$ is not diagonal. Particularly, any finite difference method applied to approximate the second derivative, cf.~Eq.~(\ref{tmat}), must be carried out with great care, since the basis functions $\chi_{m}^{i}(x)$ given in FE-DVR representation are continuous but do not have continuous derivatives at $x^i$. Here, we follow the derivation of Refs.~\cite{rescigno00}~and~\cite{scrinzi93and95} and obtain the block diagonal structure\cite{schneider06} of the kinetic-energy matrix as
\begin{widetext}
 \begin{align}
 t_{m_1m_2}^{i_1i_2}&=-\frac{1}{2}\int_{0}^{x_0}\!\!\!\dint{}{x}\,\chi_{m_1}^{i_1}(x)\,\nabla^2\,\chi_{m_2}^{i_2}(x)&\nonumber\\
\label{tmat}
&=\left\{
\begin{array}{cc}
 \frac{1}{2}\,\delta_{i_1i_2}\,\tilde{t}_{m_1m_2}^{i_1}\left[w_{m_1}^{i_1}w_{m_2}^{i_1}\right]^{-1/2}\; &\;\;\;\;,\; m_1>0,\,m_2>0\\
&\\
\frac{1}{2}\left(\delta_{i_1i_2}\,\tilde{t}_{n_g-1,m_2}^{i_1}+\delta_{i_1i_2-1}\,\tilde{t}_{0m_2}^{i_2}\right)\left[w_{n_g-1}^{i_1}+w_{0}^{i_1+1}\right]^{-1/2}\;&\;\;\;\;,\; m_1=0,\,m_2>0\\
&\\
\frac{1}{2}\left(\delta_{i_1i_2}\,\tilde{t}_{m_1n_g-1}^{i_1}+\delta_{i_1i_2+1}\,\tilde{t}_{m_10}^{i_1}\right)\left[w_{m_1}^{i_1}\left(w_{n_g-1}^{i_2}+w_{0}^{i_2+1}\right)\right]^{-1/2}\; &\;\;\;\;,\; m_1>0,\,m_2=0\\
&\\
\displaystyle\frac{\delta_{i_1i_2}\left(\tilde{t}_{n_g-1,n_g-1}^{i_1}+\tilde{t}_{00}^{i_1+1}\right)+\delta_{i_1i_2-1}\,\tilde{t}_{0, n_g-1}^{i_2}+\delta_{i_1i_2+1}\,\tilde{t}_{n_g-1,0}^{i_1}}{2\left[\left(w_{n_g-1}^{i_1}+w_{0}^{i_1+1}\right)\left(w_{n_g-1}^{i_2}+w_{0}^{i_2+1}\right)\right]^{1/2}}\; &,\; m_1=m_2=0
\end{array}
\right.\;,
\end{align}
\end{widetext}
where the quantity $\tilde{t}^{i}_{m_1m_2}$ is connected to the first derivative\cite{rescigno00} of the Lobatto shape functions via
\begin{align}
 \tilde{t}^{i}_{m_1m_2}=\sum_{m}\frac{\mathrm{d} f_{m_1}^{i}(x^{i}_{m})}{\mathrm{d}x}\,\frac{\mathrm{d} f_{m_2}^{i}(x^{i}_{m})}{\mathrm{d}x}\,w_{m}^{i}\;.
\end{align}
Eqs.~(\ref{vmat}) and (\ref{tmat}) embody analytic formulas for the kinetic and potential energy when a (finite-element) DVR basis is involved.

The most attractive feature of the FE-DVR representation is, that it can also be used together with the GGL quadrature to construct the matrix elements of the interaction operator $\hat{u}$ which is non-local \textit{and} of two-particle type. In general, the binary-interaction matrix elements (two-electron integrals) are carrying a set of four index-pairs $(i,m)$ accounting for the two-particle character of the pair interaction, see first line of Eq.~(\ref{umat}). However, in FE-DVR, using the separable form of the discretized interaction potential $u(|x-x'|)$, see Eq.~(\ref{umatsep}), we arrive at a very simple, semi-analytic expression for these matrix elements. This opens the way towards efficient NEGF calculations:
\begin{widetext}
\begin{align}
\label{umat}
 u_{m_1m_2m_3m_4}^{i_1i_2i_3i_4}&=\int_{0}^{x_0}\!\!\!\dint{}{x}\!\int_{0}^{x_0}\!\!\!\dint{}{x'}\,\chi_{m_1}^{i_1}(x)\,\chi_{m_3}^{i_3}(x')\,u(|x-x'|)\,\chi_{m_2}^{i_2}(x)\,\chi_{m_4}^{i_4}(x')\nonumber&\\
 &=\delta_{i_1i_2}\delta_{i_3i_4}\delta_{m_1m_2}\delta_{m_3m_4}\tilde{u}_{m_1m_2}^{i_1i_2}\;,
\end{align}
\end{widetext}
with the remaining, full (kernel) matrix
\begin{align}
\label{umatred}
 \tilde{u}_{m_1m_2}^{i_1i_2}=\sum_{i_3m_3} \alpha_{m_3}^{i_3} \beta^{i_1i_3}_{m_1m_3} \beta^{i_2i_3}_{m_2m_3}\;,
\end{align}
being symmetric and of dimension $n_b\times n_b$. Here, the quantities $\alpha_{m}^{i}$ are the eigenvalues of the real matrix
\begin{align}
\label{umatsep}
U_{(im)(i'm')}=u(|x_{m}^{i}-x_{m'}^{i'}|)=\sum_{i_3m_3}\alpha_{m_3}^{i_3}\tilde{\beta}_{i_3i}^{m_3m}\tilde{\beta}_{i_3i'}^{m_3m'}\;,
\end{align}
and $\beta_{mm'}^{ii'}$
are connected to the eigenvectors $\tilde{\beta}_{mm'}^{ii'}$ via
\begin{align}
 \beta_{mm'}^{ii'}=
\left\{
\begin{array}{cc}
\tilde{\beta}_{m'm}^{i'i} &,\;m>0\\
&\\
\displaystyle\frac{\tilde{\beta}_{m' (n_g-1)}^{i'i} w_{n_g-1}^{i}+\tilde{\beta}_{m' 0}^{i'(i+1)} w_{0}^{i+1}}{w_{n_g-1}^{i}+w_{i+1}^{0}} &,\;m=0
\end{array}
\right.\;.
\end{align}
The key point is that the full rank representation (\ref{umatsep}) enables us to factorize the two integrations in Eq.~(\ref{umat}) so that each integral can be separately performed by the use of the GGL quadrature. In turn, the evaluation of the two-electron integrals $u_{m_1m_2m_3m_4}^{i_1i_2i_3i_4}$ reduces to the computation of a simple matrix of dimension $n_b\times n_b$, cf.~Eq.~(\ref{umatred}).

In summary, the effort of constructing the two-electron integral in FE-DVR representation becomes comparable to computing any single-electron matrix element (such as the kinetic or potential energy) besides an additional but numerically elementary matrix diagonalization. Moreover, Eq.~(\ref{umat}) is not only memory-friendly [the required memory scales with ${\cal O}(n_b^2)$ instead of ${\cal O}(n_b^4)$!] but also permits a much more efficient evaluation of interaction contributions, especially, self-energy diagrams, see Sec.~\ref{subsec:skkbe}. This is due to the high degree of diagonality determined by the product of Kronecker deltas in Eq.~(\ref{umat}). Also, it is favorable that the integrals do not depend on the explicit form of the pair interaction.

\subsection{\label{subsec:skkbe}Schwinger/Keldysh/Kadanoff-Baym equations}
The FE-DVR basis, as set up in Sec.~\ref{subsec:basisconstruction}, allows us to expand the one-particle nonequilibrium Green's function $G(1,2)=-\mathrm{i}\,\langle \hat{T}_{\cal C}\,{\psi}(1){\psi}^\dagger(1')\rangle$\cite{gfdefinition}, with space-time arguments $1=(x,t)$, $1'=(x',t')$ and spin omitted, as
\begin{align}
\label{gf}
 G(1,1')=\sum_{i_1m_1}\sum_{i_2m_2}\chi_{m_1}^{i_1}(x)\,\chi_{m_2}^{i_2}(x')\,g_{m_1m_2}^{i_1i_2}(t,t')\;.
\end{align}
The time-dependent coefficients $g_{m_1m_2}^{i_1i_2}(t,t')$ are in general complex and vary on the complex Keldysh time-contour\cite{keldysh64} ${\cal C}$. Further, $G(1,1')$ obeys the SKKBE\cite{martin59,keldysh64,kadanoff62}
\begin{align}
\label{skkbe}
 \{\mathrm{i}\,\partial_{t}-H(1)\}\,G(1,1')&=\delta_{\cal C}(1-1')&\\
&\hspace{0.75pc} +\!\int_{\cal C}\dint{}{2}\,\Sigma[G](1,2)\,G(2,1')\;,\nonumber
\end{align}
where $H$ ($\Sigma[G]$) denotes the one-particle energy (self-energy), the time-integral is performed over $\cal C$, and Eq.~(\ref{skkbe}) is accompanied by its adjoint equation for the second time argument. Using Eq.~(\ref{gf}), the SKKBEs transform into equations of motion for the matrix $g$ [dimension is $n_b\times n_b$ with $n_b$ as defined in Eq.~(\ref{basisdim})] and attain matrix form, where $H$, $G$ and $\Sigma$ are to be replaced by their matrix components,
\begin{align}
\label{replG}
G(1,2)\;\rightarrow&\;\;g_{m_1m_2}^{i_1i_2}(t,t')\;,&\\
\label{replH}
H(1)\;\rightarrow &\;\;h_{m_1m_2}^{i_1i_2}(t)&\\
&\;=\,t_{m_1m_2}^{i_1i_2}+v_{m_1m_2}^{i_1i_2}(t)\;,&\nonumber\\
\label{replS}
\Sigma[G](1,2)\;\;\rightarrow&\;\Sigma_{m_1m_2}^{i_1i_2}[g](t,t')&\\
&\;=\,\Sigma_{m_1m_2}^{\mathrm{HF},i_1i_2}(t,t')+\Sigma_{m_1m_2}^{\mathrm{corr},i_1i_2}(t,t')\;,\nonumber
\end{align}
and all products are to be understood as matrix products. In Eq.~(\ref{replH}), $h_{m_1m_2}^{i_1i_2}(t)$ has the block-diagonal structure imprinted by the kinetic energy, cf.~Eq.~(\ref{tmat}). Moreover, Eq.~(\ref{replS}) separates the self-energy $\Sigma[g]_{m_1m_2}^{i_1i_2}(t,t')$ into Hartree-Fock (HF) and correlation parts, both of which are, generally, full [of dimension $n_b\times n_b$] and functionals of $g$. The Hartree-Fock self-energy $\Sigma^\mathrm{HF}$ and the correlation self-energy $\Sigma^\mathrm{corr}$ in second Born approximation attain the form
\begin{widetext}
\begin{align}
\label{sigmahf}
 \Sigma^{\mathrm{HF},i_1i_2}_{m_1m_2}(t,t')&=-\mathrm{i}\,\delta_{\cal C}(t-t')\left\{\sigma\,\delta_{i_1i_2}\delta_{m_1m_2}\sum_{i_3m_3}\tilde{u}_{m_1m_3}^{i_1i_3}\,g_{m_3m_3}^{i_3i_3}(t,t^+)-\tilde{u}_{m_1m_2}^{i_1i_2}\,g_{m_2m_1}^{i_2i_1}(t,t^+)\right\}\;,&\\
\label{sigma2ndb}
 \Sigma^{\mathrm{corr},i_1i_2}_{m_1m_2}(t,t')&=\sum_{i_3m_3}\sum_{i_4m_4}\left\{\sigma\,g^{i_1i_2}_{m_1m_2}(t,t')\,g^{i_3i_4}_{m_3m_4}(t,t')-g^{i_1i_4}_{m_1m_4}(t,t')\,g^{i_3i_2}_{m_3m_2}(t,t')\right\}g^{i_4i_3}_{m_4m_3}(t',t)\,\tilde{u}_{m_1m_3}^{i_1i_3}\,\tilde{u}_{m_2m_4}^{i_2i_4}\;,
\end{align}
\end{widetext}
where $\sigma\in\{1,2\}$ accounts for the spin-degeneracy, and $t^+$ indicates the limit $t\rightarrow t+\epsilon_{>0}$ from above on the contour ${\cal C}$. Equilibrium initial correlations concerning $\Sigma^{\mathrm{corr}}$ are treated in the mixed Green's function approach\cite{dahlen05,dahlen06,stan09}, where $G$ and $\Sigma$ have complex time-arguments $t_{\geq 0}+\mathrm{i}\,\bar{t}$ with $\bar{t}\in[-\beta,0]$ and $\beta$ being the inverse temperature---for the full set of equations involved see e.g.~Ref.~\cite{stan09}.

The self-energy expressions (\ref{sigmahf}) and (\ref{sigma2ndb}) manifest very simple forms which arise from the subtle structure of the FE-DVR basis, compare with Refs.~\cite{balzer09_prb,dahlen07_prl}. In the time-local HF part, Eq.~(\ref{sigmahf}), the  Hartree term is completely diagonal [just as $v$ in Eq.~(\ref{vmat})] requiring a single sum over the index pair $(i_3,m_3)$, and the exchange term involves only a product of two matrix elements. Note, that simultaneous summations over $i$ and $m$ are equivalent to a single sum with $n_b$ elements! With this in mind, the evaluation of the second Born self-energy, scales with ${\cal O}(n_b^2)$ implying only two summations per matrix element. This has to be compared with the general basis representation: there, two sums are required for each full vertex point in the second-order diagrams and, additionally, a single sum is needed for the start- as well as for the end-point leading to an effort of ${\cal O}(n_b^6)$ in total for second-order self-energies. The simplification of this is a main result of the present paper and provides the basis for addressing new classes of problems, in particular, laser-atom interactions.

In conclusion, using the FE-DVR representation in combination with the two-electron integrals $w_{m_1m_2m_3m_4}^{i_1i_2i_3i_4}$ of Sec.~\ref{subsec:matrixelements}, it is possible to rigorously reduce the computational complexity for inhomogeneous NEGF applications, at least, in one spatial dimension. In particular, with Eq.~(\ref{sigma2ndb}), the effort becomes comparable to that in lattice models, see~e.g.~\cite{myohanen08,thygesen08,vanfriesen09}, which, by construction, are computationally much simpler. Once the Green's function $G(1,1')$ is computed from the matrix form of Eq.~(\ref{skkbe}), many observables such as the one-electron density $n(x,t)=-\mathrm{i}\,G(1,1^+)$, the time-dependent dipole moment (and in turn the polarizability\cite{dahlen07_prl}) or the total energy are accessible\cite{evaluation}.

\section{\label{sec:results}Model atoms and molecules}
In this section, we apply the FE-DVR representation, Eq.~(\ref{gf}), to compute the nonequilibrium Green's function for atomic and molecular few-electron model systems. As atomic example, we discuss the one-dimensional helium atom (1D He), e.g.~\cite{pindzola91,grobe93,bauer97,liu99,dahlen01}, which represents the most elementary closed-shell system. This model of the 3D helium atom has been studied since the 1970s and is known to reliably provide the qualitative features of the single- and double-ionization dynamics in intense laser fields\cite{lein00} including the knee structure\cite{dahlen01}. Moreover, it is still actively considered, e.g.~\cite{zanghellini04,ruggenthaler09}, as it serves as a fundamental 'testing ground' for multi-electron calculations. This issue is due to the presence of strong electron-electron ($e$-$e$) correlations which require a treatment beyond mean-field (HF) theories. In addition to He, we discuss two molecular models with two and four electrons, respectively: The hydrogen molecule (H$_2$), e.g.~\cite{lapidus82,yu96,kawata00,baier07,christov08}, and lithium hydride (LiH)\cite{tempel09}---again in one spatial dimension. The reason why we focus on these atomic and molecular systems is twofold: (i)~the long-range character of the ionic Coulomb potential (enhanced in 1D!) proves the vital necessity for extended basis sets for the construction of which the FE-DVR is indeed well suitable, and (ii)~the possible comparison to exact solutions, obtained from the time-dependent Schr\"odinger equation (TDSE), allows us to verify the quality of the involved many-body approximations. Also, in the present paper, we restrict the NEGF calculations to the ground states.

\begin{table}[t]
\begin{tabular*}{1.0\columnwidth}{@{\extracolsep{\fill}} l l l l}
\hline
\hline
&&&\\
\textbf{Hartree-Fock}&  & $n_g$ [$n_b$] & $E_\mathrm{gs}^\mathrm{HF}$ [a.u.] \\
\hline
 &  & $4$ [$43$] & $-2.22\cdot\cdot\cdot\cdot\,\cdot$ \\
 &  & $9$ [$98$] & $-2.224209\,\cdot$ \\
 &  & $14$ [$153$] & $-2.2242096$ \\
&&&\\
\textbf{Second Born} & $n_g$ [$n_b$] & \# $\tau$-grid points & $E_\mathrm{gs}^\mathrm{2ndB}$ [a.u.] \\
\hline
 & $14$ [$153$] & $101$ & $-2.23\cdot\cdot\cdot\cdot\,\cdot$ \\
 & $14$ [$153$] & $301$ & $-2.2334\cdot\cdot\,\cdot$ \\
 & $14$ [$153$] & $601$ & $-2.23341\cdot\cdot$ \\
 & $14$ [$153$] & $1001$ & $-2.233418\,\cdot$ \\
&&&\\
\textbf{TDSE (exact)} &  &  & $E_\mathrm{gs}^\mathrm{TDSE}$ [a.u.] \\
\hline
 &  &  & $-2.2382578$ \\
&&&\\
\hline
\hline
\end{tabular*}
\caption{Ground state energy $E_\mathrm{gs}$ of the 1D He atom (with fully converged decimal places) as computed from the Green's function in Hartree-Fock and second Born approximation. The exact energy is obtained from the time-dependent Schr\"odinger equation (TDSE). $153$ FE-DVR basis functions [at $n_e=11$] are adequate to reach the HF-limit and, thus, convergence with respect to the basis size. In second Born approximation, about $600$ points in imaginary time are needed for convergence in the fifth decimal place.}\label{tab1}
\end{table}

\begin{figure}[t]
\includegraphics[width=0.455\textwidth]{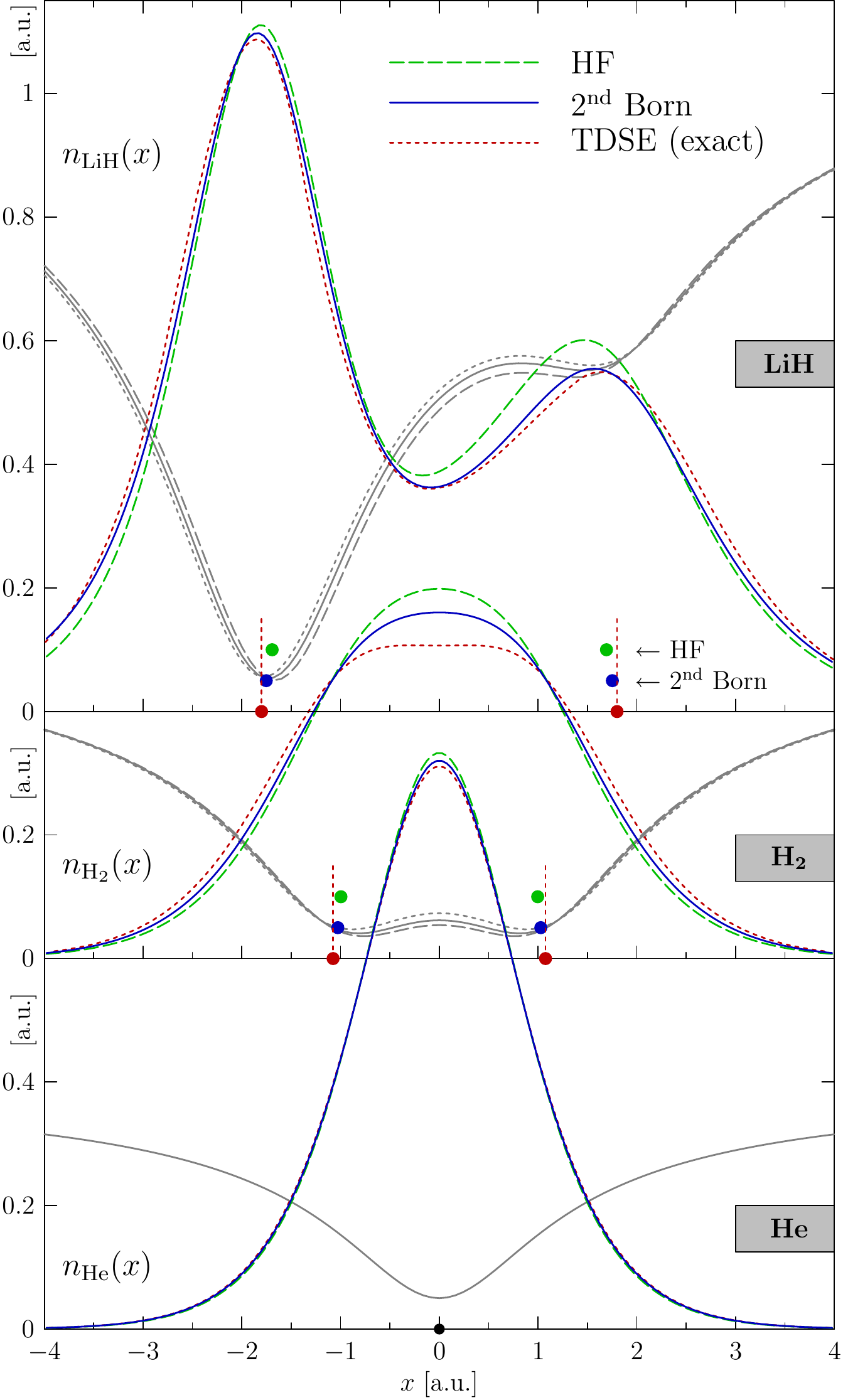}
 \caption{(Color online) One-electron ground-state density $n(x)$ of the one-dimensional He atom, the H$_2$ and the LiH molecule in Hartree-Fock (dashed) and second Born approximation (solid). The exact density (dotted) is obtained from the time-dependent Schr\"odinger equation (TDSE) with imaginary time propagation. The colored dots show the equilibrium positions of the ions separated by the bond-length $d_\mathrm{b}$, Table~\ref{tab2}, and the gray curves indicate the associated potentials $v(x)$ [in a.u.~(left ordinate) but scaled by factor $0.35$ and shifted].}\label{fig2}
\end{figure}

In hamiltonian (\ref{ham}), the helium atom is modeled by using $v(x)=-Z[(x-x_0/2)^2+1]^{-1/2}$ as regularized potential, where the atomic number is $Z=2$. Thereby, the $x_0/2$-shift ensures that the nucleus is situated in the center of the discretized interval $[0,x_0]$. For the hydrogen molecule and lithium hydride the coordinate is taken along the bond axis such that the potential is given by $v_d(x)=-Z_1 [(x-(x_0+d)/2)^2+1]^{-1/2}-Z_2[(x-(x_0-d)/2)^2+1]^{-1/2}$, where $d$ denotes the interatomic distance, $Z_1$=1 and $Z_2=1$ for the hydrogen and $Z_2=3$ for the lithium atom. Principally, the regularization parameters [here, $1$ for H and Li] can be adjusted to match the difference of the ionization potentials of the individual model atoms to the 3D atoms, see Ref.~\cite{tempel09}. Furthermore, for all three systems, a soft-core Coulombic $e$-$e$ pair potential has been applied: $u(|x-x'|)=\left[(x-x')^2+1\right]^{-1/2}$.

For the 1D helium atom, we have used $11$ finite-elements within an interval of $x_0=50$ a.u.~length. Some smaller FEs have, thereby, been arranged around $x_0/2$ to ascertain larger numerical precision in the central region. Further, the number of local DVR basis functions $n_g+1$ has been varied between $5$ and $20$ to obtain convergence of the ground-state energy $E_\mathrm{gs}$, and, in Eqs.~(\ref{sigmahf}) and (\ref{sigma2ndb}), the spin-degeneracy factor was set to $\sigma=2$ leading to the singlet state.

For the Hartree-Fock approximation, the convergence of the He ground-state energy---at the fixed FE-configuration---is shown in Table~\ref{tab1} with regard to the basis size.
At $n_g=14$ , corresponding to $153$ basis functions in total, we obtain the HF-limit with more than six decimal places precision and, consequently, sufficient convergence with respect to the basis dimension. For the second Born approximation, we used the same FE-DVR set-up. However, due to the grand-canonical averaging involved in $G(1,1')$, see definition in Sec.~\ref{subsec:skkbe}, the ground-state [equilibrium] Green's function has an additional imaginary time argument $\tau=t-t'\in[-\mathrm{i}\beta,0]\subset{\cal C}$. This has been discretized using a uniform power mesh, for details see e.g.~Refs.~\cite{dahlen05,balzer09_prb}, and to ensure the zero-temperature limit, i.e.~the ground state, we set $\beta=100$. We note that, in the HF case, this grid is redundant as $\Sigma^\mathrm{HF}(t,t')$ is local in time, cf.~Eq.~(\ref{sigmahf}). For the second Born calculation, this implies, though, checking convergence with respect to a second parameter: the number of $\tau$-grid points, see Table~\ref{tab1}. In $2^\mathrm{nd}$ Born approximation, the helium ground-state energy converges towards $-2.2334$~a.u., which is $0.0092$~a.u.~lower than the HF reference value, and a comparable accuracy is obtained by using more than $600$ time-grid points. With a deviation of less than $0.005$~a.u., it comes close to the exact ground state\cite{he_gs_energy} ($-2.2383$~a.u.), which follows from the TDSE.

\begin{figure}[t]
\includegraphics[width=0.435\textwidth]{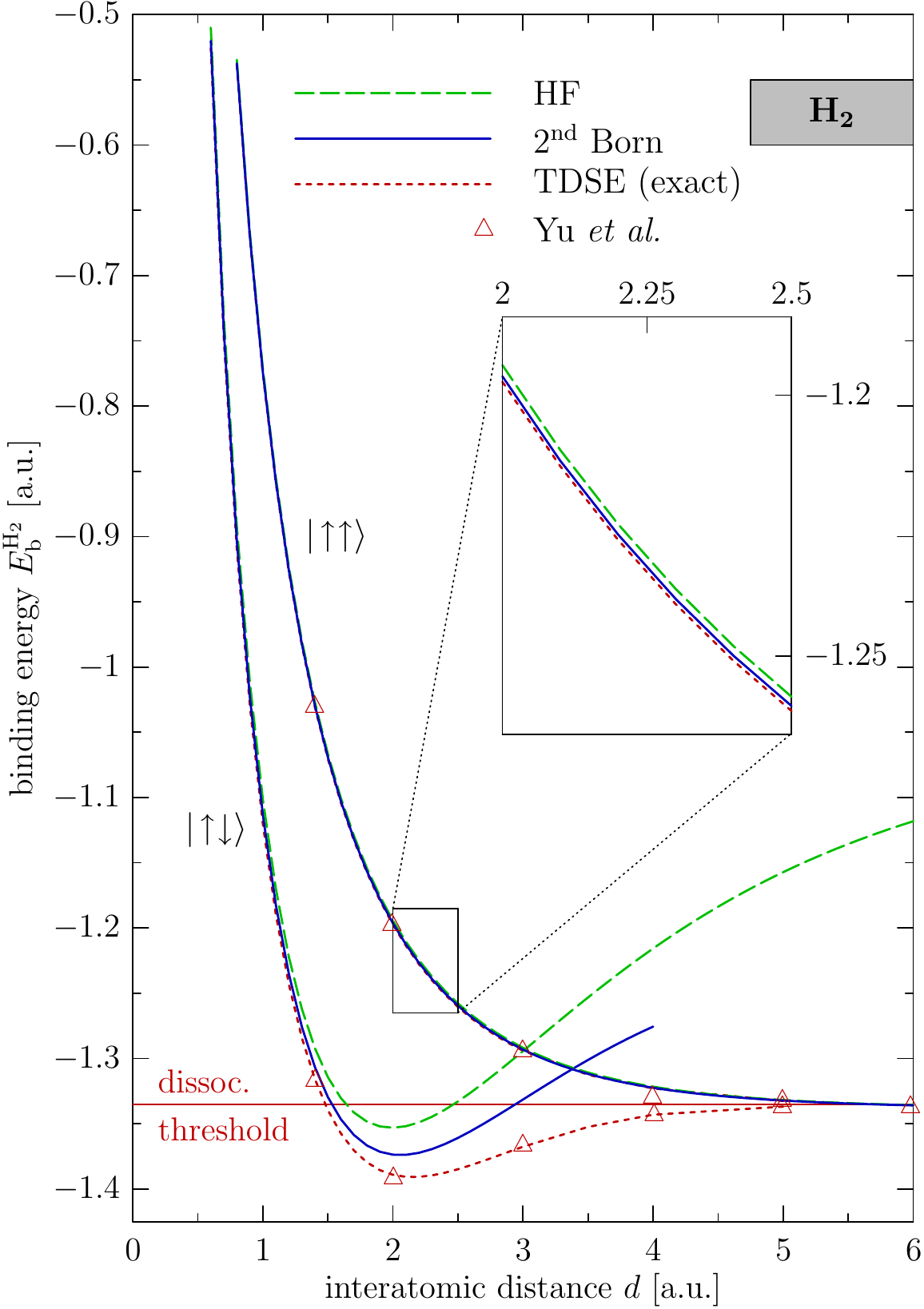}
 \caption{(Color online) Hydrogen-hydrogen binding energy $E_\mathrm{b}^{\mathrm{H}_2}$ as function of the interatomic distance $d$ for the case of the binding singlet ($|\!\!\uparrow\downarrow\rangle$) and anti-binding triplet, spin-polarized ($|\!\!\uparrow\uparrow\rangle$) system. While the triplet system is less affected by correlations (see inset figure), the binding-energy curve of the singlet state is essentially improved against HF in the $2^\mathrm{nd}$ Born approximation. The triangles mark TDSE results from Ref.~\cite{yu96}. The exact dissociation threshold is indicated by the horizontal line. For the values of the bond-lengths see Table~\ref{tab2}.
}\label{fig3}
\end{figure}

The one-electron ground-state density for the 1D He atom is obtained from $n(x)=-\mathrm{i}\,G(x, t;x, t')|_{\tau\rightarrow 0+\mathrm{i}0^-}$ and is displayed in the bottom graph of Fig.~\ref{fig2}. The differences of both approximate results (dashed/solid line) and the exact density (dotted line) are most dominant within a small range of $0.5$ a.u.~around the ion position. As for the total energies, the second Born density improves the HF result and is relatively close to the exact density profile.

\begin{table}[t]
\begin{tabular*}{1.0\columnwidth}{@{\extracolsep{\fill}} l l l l l}
\hline
\hline
&&&&\\
Bond-length $d_\mathrm{b}$ &  & \textbf{HF} & $2^\mathrm{nd}$ \textbf{Born} & \textbf{exact}\\
\hline
   &&&&\\
  &H$_2$ & $1.9925$ & $2.0561$ & $2.151$\\
  &LiH & $3.3860$ & $3.5053$ & $3.6\cdot\cdot$\\
Binding energy $E_\mathrm{b}$ &  & \textbf{HF} & $2^\mathrm{nd}$ \textbf{Born} & \textbf{exact}\\
\hline
   &&&&\\
  &H$_2$ & $-1.3531$ & $-1.3740$ & $-1.391$\\
  &LiH & $-4.8534$ & $-4.8886$ & $-4.91\,\cdot$\\
&&&&\\
\hline
\hline
\end{tabular*}
\caption{Computed equilibrium bond-lengths $d_\mathrm{b}$ and corresponding binding energies $E_\mathrm{gs}(d_\mathrm{b})+Z_1Z_2/d_\mathrm{b}$ of the one-dimensional H$_2$ and LiH model [all quantities in a.u.]. While the Hartree-Fock and second Born values are Green's function results, the exact values are obtained from the full solution of the few-particle TDSE.}\label{tab2}
\end{table}

\begin{figure}[t]
\includegraphics[width=0.435\textwidth]{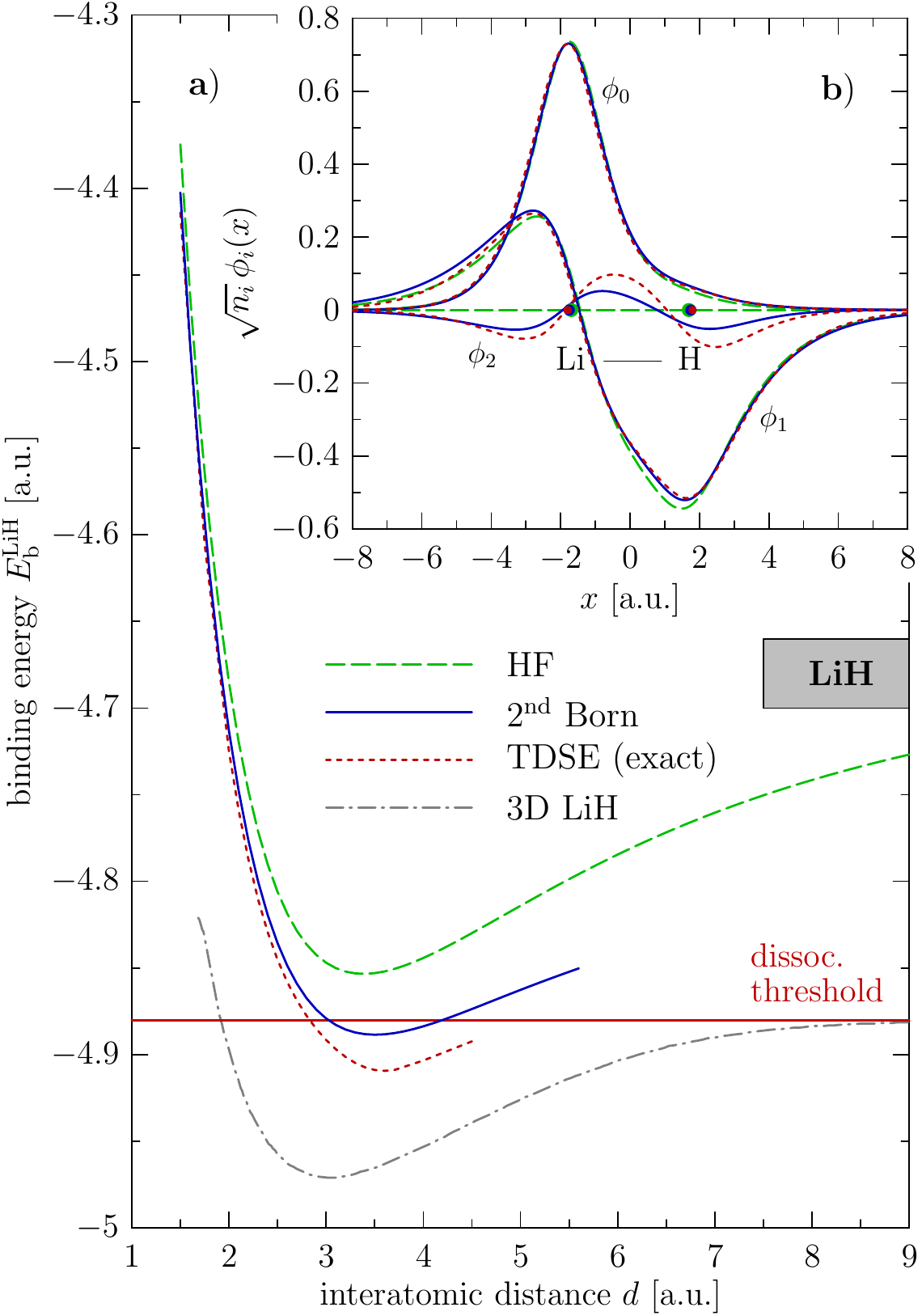}
 \caption{(Color online) a) Li-H binding energy $E_\mathrm{b}^{\mathrm{LiH}}$ as function of the interatomic distance $d$. For the specific bond-lengths see Table~\ref{tab2}. The compound dissociates at a threshold of $\ldots$~a.u.. For comparison, the dash-dotted line shows the binding-energy curve for the three-dimensional molecule\cite{juarros06}. b) Most relevant natural orbitals $\phi_i(x)$ for the LiH ground state at equilibrium bond-length as obtained from the TDSE, the HF and second Born Green's function [weighted by their occupation $n_i$].}\label{fig4}
\end{figure}

For the hydrogenic and the lithium hydride system, the electron ground-state energy changes with distance $d$ between the atomic nuclei. Hence, whether [or whether not] the individual atoms combine into molecules, depends on the H-H (Li-H) binding energy $E_\mathrm{b}(d)$ which is the electron ground-state energy plus the interatomic repulsion\cite{interat.rep} $(Z_1 Z_2)/d$. The computed binding-energy curves for H$_2$ are displayed in Fig.~\ref{fig3}, where a FE-DVR set-up similar to the helium case has led to convergent results. The singlet state $|\!\!\uparrow\downarrow\rangle$, again with $\sigma=2$ in the self-energies, is binding showing a minimum at a distance $d_\mathrm{b}$ in the exact result (dotted curve) and a well defined dissociation threshold (horizontal line). Also, the HF (dashed line) and the second Born approximation (solid line) confirm a substantial hydrogen-hydrogen binding, where $2^\mathrm{nd}$-Born correlations lead to a larger binding energy that indicates an essential improvement of about $60$\% in the HF energy discrepancy. However, for the singlet state, neither the Hartree-Fock nor the second Born approximation can accurately resolve the dissociation threshold at $-1.3396$~a.u.. This is due to the fact that the closed-shell H$_2$ molecule dissociates into open-shell fragments---single hydrogen atoms. Such a transition can not be described within the semi-local (spin-restricted) approximations involved in the NEGF. We note that, the same problem is encountered in time-dependent density functional theory (TDDFT) using exact exchange\cite{ruzsinsky06}. Nevertheless, the different equilibrium positions of the nuclei (bond-lengths $d_\mathrm{b}$) and the ground-state energies are not affected by this failure of the many-body approximations, see Table~\ref{tab2}. Overall, it turns out that correlations cause larger bond-lengths due to the lower electronic ground-state energy.

We, in addition, have performed calculations for the spin-polarized or triplet H$_2$ system, $|\!\!\uparrow\uparrow\rangle$ with $\sigma=1$. The respective binding-energy curves in Fig.~\ref{fig3} show that, in contrast to the singlet state, as expected, it does not undergo molecular binding but behaves correctly in the limit $d\rightarrow\infty$. In particular, for all interatomic distances, the exact binding energy is well approximated within HF. Correlations generally improve the results (see inset of Fig.~\ref{fig3}) but play a minor role. This is typical for spin-polarized systems.

The approximate and exact one-electron ground-state density for the H$_2$ singlet is shown in Fig.~\ref{fig2} (center graph) with respect to the corresponding equilibrium bond-lengths. Thereby, the gray curves illustrate the ionic potentials $v_{d_{\mathrm{b}}}(x)$ on the bond axis. The exact density profile indicates onset of electron localization on the individual hydrogen atoms. This is not captured in HF and $2^\mathrm{nd}$ Born approximation, which both lead to a smooth profile, but the trend towards a lower density between the nuclei becomes obvious. In particular, the ionic potential with the second-Born value of $d_\mathrm{b}$ [compare also the dots in Fig.~\ref{fig2}] is in good agreement with the TDSE result.

The four-electron molecule, lithium hydride, serves as a simple example for the hetero-atomic dissociation. However, LiH, just like molecular hydrogen, dissociates into open-shell components---Li($3e$) and H($1e$). Thus, in Fig.~\ref{fig4}~a), we obtain a similar behavior of $E_\mathrm{b}^\mathrm{LiH}$ in HF and $2^{\mathrm{nd}}$ Born approximation compared to H$_2$ against interatomic distance. Within the calculations, we used a basis consisting of $15$ non-equidistant elements and up to $15$ local DVR basis functions for large values of $d$. For LiH, the inclusion of $e$-$e$ correlations, improves the results such that the minimum in the binding energy becomes situated below the exact dissociation threshold. This is not realized in HF approximation. For reference, we, in Fig.~\ref{fig4}~a), also included the binding-energy curve  for the three-dimensional counterpart\cite{juarros06} (dash-dotted line) of the 1D model, which possesses a stronger Li-H bond at comparable internulear distance. However, we note, that bond-lengths and binding energies are very  sensitive to the softening parameters used in the ion and Coulomb potentials, e.g.~Ref.~\cite{tempel09}. For the specific values of $E_\mathrm{b}$ and $d_b$ for lithium hydride, see Table~\ref{tab2}. The one-electron density of LiH is plotted in the top graph of Fig.~\ref{fig2}, where the lithium (hydrogen) atom is situated at negative (positive) $x$-positions, cf.~the ion potentials $v_d(x)$. In all considered cases, the density shows a clear minimum between the nuclei, and correlations mainly alter the density around the hydrogen atom. In particular, we highlight, that the second Born ground-state density is in surprisingly good agreement with the exact result.

In order to identify the intra-molecular electronic structure more closely, we, in addition, have computed the most relevant natural orbitals (NO) for the 1D LiH molecule, see Fig.~\ref{fig4}~b). The natural orbitals $\phi_i(x)$, $i=0,1,\ldots,n_b-1$, are obtained from the eigenvalue problem
\begin{align}
\int\dint{}{x'} \rho(x,x')\phi_i(x')=n_i\phi_i(x)\;, 
\end{align}
with density matrix $\rho(x,x')=-\mathrm{i}\,G(1,1')_{\tau\rightarrow0+\mathrm{i}0^-}$ and occupations $n_i\in[0,1]$.
For HF ground states, with $\beta\rightarrow\infty$, we have $n_i=1$ for $i=0,1,\ldots,N-1$ and zero otherwise, where $N$ is the number of electrons with the same spin projection. Hence, there are two fully occupied orbitals for the case of lithium hydride in Hartree-Fock approximation, see the NO $\phi_0$ and $\phi_1$ in Fig.~\ref{fig4}~b). Correlation effects generally lead to occupations of more than two orbitals, cf.~$\phi_2$ in second Born approximation, compare with the exact result, and note that the orbitals have been scaled by $\sqrt{n_i}$. On the contrary, the two core electrons at the lithium atom, according to the localized NO $\phi_0$, are very little affected by correlations, which is revealed by the \mbox{HF-,} $2^\mathrm{nd}$-Born- and the TDSE-curves lying almost on top of each other. Further, the second natural orbital $\phi_1$---with about $95$-$98$\% occupation and a node near the lithium atom---is shared between the nuclei and extends over several bond-lengths. In correspondence to the one-electron density in Fig.~\ref{fig2}, the exact $\phi_1$ is well approached by the second Born approximation, which shows the correct trend in the central bond region. Also, the third NOs $\phi_2$ are similar in shape. However, the deviation in their occupations, is mainly responsible for the differences of the $2^\mathrm{nd}$ Born to the exact result. Finally, all other natural orbitals (those which are not shown) are occupied by less than $1$\%.

\section{\label{sec:conclusion}Conclusion}
In this work, we have applied the finite-element discrete variable representation (FE-DVR) to expand the one-particle nonequilibrium Green's function with respect to the two [one-dimensional] spatial coordinates. This procedure is highly favorable against a general basis representation: (i)~conceptionally, it allows for an optimal and flexible combination of grid \textit{and} basis methods, (ii),~with respect to the NEGF of finite systems, a direct solution of the SKKBE within a grid-based hybrid approach becomes possible by (iii) an essentially simplified treatment of all binary interactions. The latter point includes the description of particle-particle correlations, where the second-order Born self-energy in Sec.~\ref{subsec:skkbe}, as the most basic model of correlations, attains a comparably simple form induced by the high degree of diagonality involved in the two-electron integrals, Eq.~(\ref{umat}), expressed in the FE-DVR picture. Also, due to the discretization in coordinate space, it is straightforward to change the one-particle potential $v(x)$ or the specific form of the pair interaction $u(|x-x'|)$. This is in striking contrast to a general basis, where to some extent enormous, extra numerical effort is required if the matrix elements and/or two-electron integrals are not analytically accessible and have to be precomputed. This, completely, drops out in the present approach.

In summary, the developed method enables better performance with relation to larger accuracy and spatial resolution, but at crucially lower numerical cost---regarding storage memory \textit{and} computing time. In particular, this also holds true when spatially extended hamiltonians are being considered, as shown in Sec.~\ref{sec:results}. In turn, applying the FE-DVR, larger basis dimensions with a guide number of $n_b\approx500-1000$ become feasible, which implies an enhancement of more than one order of magnitude compared to a general basis approach.

For illustration purposes, we have computed the nonequilibrium Green's function for simple but benchmarking atomic and molecular models: The helium atom and the linear molecules H$_2$ and LiH in one spatial dimension. Especially for the molecular systems, where two (four) electrons are shared between the nuclei, the enhanced electron collision rate in one dimension makes it attractive to investigate electron-electron correlation effects in second Born approximation. Indeed, with respect to inhomogeneous and finite systems, only few comparisons of NEGF findings to exact many-body results are available. In our comparisons, we restricted ourselves to two- and four-electron models as the full solution of the TDSE becomes impractical for more than four electrons. In the present examples, it turns out, that the $2^\mathrm{nd}$ Born approximation is well capable to catch the main ground-state features of the considered models. Thus, the presented analysis affirmatively contributes to the assessment of the applicability of NEGFs to atomic and molecular systems.

Of course, the FE-DVR approach enables calculations also with larger particle numbers. Depending on the system, multi-electron ensembles [in one spatial dimension] with up to $N\lesssim20$ turn out to be feasible\cite{balzer_pngf4_09}. Particularly, we note that, with this grid-based method adequate spatial resolution over a range of several hundred atomic units becomes, for the first time, available in NEGF approaches to strongly inhomogeneous quantum systems. The good performance is thereby not limited to the second Born approximation. The method also allows for more complicated self-energies including $GW$ or $T$-matrix on spatial grids. Moreover, the attractive scaling behaviors of the FE-DVR fully survive in nonequilibrium situations and, thus, provide essential impact for the efficient solution of the two-time SKKBE for atomic and molecular systems. Explicit results of the time-evolution in second Born approximation, including transitions to few-electron resonance states\cite{tanner00} located energetically above the one-electron excitations, will be the subject of a forthcoming publication.

\begin{acknowledgments}
This work was, in part, supported by the Deutsche Forschungsgemeinschaft via SFB-TR 24.
\end{acknowledgments}

\appendix*
\section{Generalized Gauss-Lobatto integration}
In numerical analysis, the generalized Gauss-Lobatto (GGL) scheme is a special quadrature rule which approximates a definite integral of a function $g(x)$ as
\begin{align}
\label{gglq}
 \int_{0}^{x_0}\!\!\!\dint{}{x}\,g(x)=\sum_{i}\int_{x^{i}}^{x^{i+1}}\!\!\!\!\!\!\dint{}{x}\,g(x)\approx\sum_i\sum_{m=0}^{n_g-1}\,g(x^{i}_{m})\,w_{m}^{i}\;,
\end{align}
where the specified points $x_m^i$ and weights $w_m^i$ are associated with sub-domains $[x^i,x^{i+1}]$ or finite elements $i$ of the integration, see definition in Sec.~\ref{subsec:basisconstruction}. For an arbitrary segmentation of the total domain $[0,x_0]$, the approximation becomes exact in the limit $n_g\rightarrow\infty$. Moreover, from the GGL integration it follows, that the Lobatto shape functions are orthogonal in the sense of the quadrature rule:
\begin{align}
 \int\dint{}{x}\,f_{m}^{i}(x)\,f_{m'}^{i'}(x)&=\delta_{ii'}\sum_{\bar{m}}f_{m}^{i}(x^{i}_{\bar{m}})\,f_{m'}^{i}(x^{i}_{\bar{m}})\,w^{i}_{\bar{m}}&\nonumber\\
 &=\delta_{ii'}\,\delta_{mm'}\,w_{m}^{i}\;.
\end{align}




\begin{thebibliography}{100}
\bibitem{gfdefinition} ${\psi}(1)$ and ${\psi}^\dagger(1')$ are electron field operators, $\hat{T}_{\cal C}$ denotes time-ordering on the contour $\cal C$, see e.g.~Ref.~\cite{keldysh64}, and $\langle\ldots\rangle$ indicates ensemble averaging.
\bibitem{evaluation} We note that, wherever integrals over coordinate space emerge in these evaluations, the generalized Gauss-Lobatto quadrature rule has to be used for consistency with Eq.~(\ref{basis}).
\bibitem{he_gs_energy} The 1D helium model can easily be adjusted to reproduce the ground-state energy of the 3D helium atom, e.g.~\cite{zanghellini04}.
\bibitem{interat.rep} The internulear repulsion
is not of soft-core type.
\end{thebibliography}
\end{document}